\documentclass[%
 reprint,
 amsmath,amssymb,
 aps,
]{revtex4-2}
\usepackage{color}
\usepackage{graphicx}% Include figure files
\usepackage{dcolumn}% Align table columns on decimal point
\usepackage{bm}% bold math
\begin{document}

\preprint{APS/123-QED}

\title{Shell effects on the drift and fluctuation in multinucleon transfer reactions}% Force line breaks with \\

\author{Zehong Liao$^{1}$}
\author{Zepeng Gao$^{1}$}%
\author{Yu Yang$^{1}$}%
%\author{Yueping Fang$^{1}$}%
 \affiliation{Sino-French Institute of Nuclear Engineering and Technology, Sun Yat-sen University, Zhuhai 519082, China}%Lines break automatically or can be forced with \\

\author{Long Zhu$^{1,2}$}
 \email{Corresponding author: zhulong@mail.sysu.edu.cn}
\author{Jun Su$^{1,2}$}
 \affiliation{$^{1}$Sino-French Institute of Nuclear Engineering and Technology, Sun Yat-sen University, Zhuhai 519082, China\\
 $^{2}$College of Physics and Technology and Guangxi Key Laboratory of Nuclear Physics and Technology, Guangxi Normal University, Guilin 541004, China}%

\date{\today}% It is always \today, today,
             %  but any date may be explicitly specified

\begin{abstract}
This study employs the dinuclear system model (DNS-sysu) to investigate drift and fluctuation mechanisms in $^{136}$Xe + $^{209}$Bi collisions above the Coulomb barrier. The DNS-sysu model demonstrated its effectiveness in providing reasonable descriptions of drift and fluctuation dynamics for the multinucleon transfer reaction at low energies. We investigate temperature-induced changes in the shell effect, which impacts nucleon transfer. At higher energies, the weakening constraint of the potential energy surface leads to a reversal in evolution direction. Additionally, the consideration of shell corrections notably affects fragment distribution at low energies but diminishes for high-energy conditions. This research provides valuable insights into understanding the macroscopic manifestation of nucleon transfer in the multinucleon transfer reaction. 
\end{abstract}

%\keywords{Suggested keywords}%Use showkeys class option if keyword
                              %display desired
\maketitle

%\tableofcontents

\section{\label{sec:level1}INTRODUCTION}
In quantum many-body systems, transport phenomena are observed in heavy-ion collisions. 
%The potential energy surface (PES) plays a crucial role in determining the evolution path of the system. 
At low energy collisions, the interplay between quantum effects and thermal dynamics shaping the nuclear reaction process becomes particularly significant \cite{simenel2020timescales}. Since the 1970s, people have discovered the phenomenon of multinucleon transfer (MNT) reactions with extensive products and strong energy dissipation \cite{volkov1978DIC}. In addition to being a very promising method for synthesizing new nuclides \cite{ WOS:000259377500026, watanabe2015pathway}, the MNT is also regarded as an ideal method for studying transport phenomena in quantum systems \cite{mcintosh2019interplay}. Some of the main underlying features of the MNT reaction can be assessed by a discussion of the drift and fluctuation mechanisms, but the interplay between the nucleon transfer process and these mechanisms in quantum systems is still not well understood. 

The problem of determining the influence of fluctuations around the macroscopic observables of MNT reaction has received much attention in recent years and has given rise to much literature \cite{PhysRevLett.105.192701, zagrebaev2011production, li2016multinucleon, zhao2021production, wu2019microscopic, zhang2024multinucleon, 2020saikomnt}. Many researchers have made notable progress in understanding MNT collisions through the use of master equation-type descriptions \cite{PhysRevC.68.034601, zhu2017theoretical, bao2018influence, wen2020dydns, PhysRevC.100.054616, zhu2021unified}. By restricting the expansion to second-order terms, W. N$\text{\"{o}}$renberg transformed the master equations into equations of Fokker-Planck type and applied it to analyze the transport phenomenon in the reaction $^{40}$Ar+$^{232}$Th successfully \cite{NORENBERG1974289}. Afterward, a considerable amount of work on extracting transport coefficients from microscopic theories was published thereafter \cite{norenberg1975quantum, ayik1976microscopic}. The findings have encouraged that it is possible to directly extract the distribution probability of fragments from the master equations using statistical treatments \cite{W.Li_2003}. This approach offers a promising avenue for gaining insights into the behavior of MNT reactions and understanding the role of fluctuations in shaping the outcomes.

As shown in Ref. \cite{Haag2017ModellingWT}, the master equation can provide a simple description of nonequilibrium systems, useful for testing the validity of fluctuation relations, and has been used to describe the diffusion processes in the collective variables characterizing the heavy-ion collision in the low energy. Based on master equations, the dinuclear system (DNS) model can provide the reasonable probability distribution of collective variables including the mass and charge yield, and can be applied to describe multiple reaction channels including quasifission, fusion, and multinucleon transfer processes. 
%Thus, master equations can be adapted to describe the behavior of the mass and charge drift and fluctuation mechanisms in the MNT reactions. 
This makes it possible to provide a reasonable average or variance value of the fragments in the MNT reactions. The purpose of this paper is to provide some useful insight into the interplay between the fluctuation of the fragment distribution and the PES based on the DNS model. The results can help in understanding how nucleons move in response to both deterministic evolution path (drift) and random distribution (fluctuation) for the MNT reaction. 

In this work, we investigate the drift and fluctuation mechanism in the collision of the asymmetric $^{136}$Xe + $^{209}$Bi system. The experimental data of $^{136}$Xe + $^{209}$Bi includes abundant PLF atomic numbers, energy, and angular distributions, as well as the correlation between these observable values \cite{SCHRODER1978301, xebi861, xebi526684}. This creates a good opportunity for theoretical model testing \cite{ Schroder1977Damped, riedel1980statistical, PhysRevC.96.024618, PhysRevC.29.1331}, and it is also an ideal reaction to understand the correlation between drift, fluctuation, and energy dissipation during the nucleon exchange process \cite{PhysRevC.27.1328, ACMerchant_1983}. In Sec. \ref{sec: model}, we present a formal description of the nucleon drift and fluctuation in the DNS-sysu model (a version of DNS models). In Sec. \ref{sec: results}, we carry out calculations of variances of fragment distributions and show the shell effect and energy dependence on the fragment distribution. Conclusions are given in Sec. \ref{sec: conclusions}.

\section{MODELS}\label{sec: model}
The DNS-sysu model has been described in detail in Ref. \cite{zhu2021unified}. Here we provide only its basics. The model has three degrees of freedom describing the macroscopic observables of projectile-like fragments (PLF). Under the framework of master equations, the joint probability of fragments $P$ evolving with different macroscopic degrees of freedom can be solved numerically as follows:
\begin{flalign}
\begin{split}\label{master}
&\frac{dP(Z_{1}, N_{1}, \beta_{2}, J, t)}{dt}\\
&=\sum_{Z_{1}^{'}}W_{Z_{1}, N_{1}, \beta_{2}; Z_{1}^{'}, N_{1}, \beta_{2}}(t)[d_{Z_{1}, N_{1}, \beta_{2}}P(Z_{1}^{'}, N_{1}, \beta_{2}, J, t)\\
&-d_{Z_{1}^{'}, N_{1}, \beta_{2}}P(Z_{1}, N_{1}, \beta_{2}, J, t)]\\
&+\sum_{N_{1}^{'}}W_{Z_{1}, N_{1}, \beta_{2};Z_{1}, N_{1}^{'}, \beta_{2}}(t)[d_{Z_{1}, N_{1}, \beta_{2}}P(Z_{1}, N_{1}^{'}, \beta_{2}, J, t)\\
&-d_{Z_{1}, N_{1}^{'}, \beta_{2}}P(Z_{1}, N_{1}, \beta_{2}, J, t)]\\
&+\sum_{\beta_{2}^{'}}W_{Z_{1}, N_{1}, \beta_{2};Z_{1}, N_{1}, \beta_{2}^{'}}(t)[d_{Z_{1}, N_{1}, \beta_{2}}P(Z_{1}, N_{1}, \beta_{2}^{'}, J, t)\\
&-d_{Z_{1}, N_{1}, \beta_{2}^{'}}P(Z_{1}, N_{1}, \beta_{2}, J, t)].
\end{split}
\end{flalign}
Here, $W_{Z_{1}, N_{1}, \beta_{2};Z_{1}^{'}, N_{1}, \beta_{2}}$ denotes the mean transition probability from the channel ($Z_{1}$, $N_{1}$, $\beta_{2}$) to ($Z_{1}^{'}$, $N_{1}$, $\beta_{2}$), which is similar to $N_{1}$ and $\beta_{2}$. $d_{Z_{1}, N_{1}, \beta_{2}}$ is the microscopic dimension (the number of channels) corresponding to the macroscopic state ($Z_{1}$, $N_{1}$, $\beta_{2}$). The expression of transition probability can be seen in Ref. \cite{PhysRevC.98.034609}. The interaction time in the dissipative process of two colliding nuclei is determined by using the deflection function method \cite{li1983distribution}.

Under collision conditions with the arbitrary angular momentum $J$, the evolution of the mean (expectation) values of the proton number $Z_{1}$ and neutron number $N_{1}$ of the PLF over time can be evaluated as: 
\begin{flalign}
\begin{split}\label{NZbar}
\bar{Z}_{1}(t)=\langle{Z}_{1}(t)\rangle=\sum_{Z_{1}}\sum_{N_{1}}\sum_{\beta_{2}}Z_{1}\times P(Z_{1}, N_{1}, \beta_{2}, t)\\
\bar{N}_{1}(t)=\langle{N}_{1}(t)\rangle=\sum_{Z_{1}}\sum_{N_{1}}\sum_{\beta_{2}}N_{1}\times P(Z_{1}, N_{1}, \beta_{2}, t).
\end{split}
\end{flalign}

%The rate of change of the proton or neutron number of PLF can be deduced as
%\begin{flalign}
%\begin{split}
%\frac{d}{dt}\bar{Z}_{1}(t)=v_{Z}(t), \qquad \frac{d}{dt}\bar{N}_{1}(t)=v_{N}(t).
%\end{split}
%\end{flalign}
%Here $v_{Z(N)}(t)$ denotes the proton(neutron) drift coefficient.
Furthermore, the variances of neutron and proton distribution with the arbitrary angular momentum $J$ are defined as:
\begin{flalign}
\begin{split}\label{sigma}
 \sigma^{2}_{NN}(t)=\sum_{Z_{1}}\sum_{N_{1}}\sum_{\beta_{2}}&(N_{1}-\bar{N}_{1}(t))^{2}\times P(Z_{1}, N_{1}, \beta_{2}, t)\\
 \sigma^{2}_{ZZ}(t)=\sum_{Z_{1}}\sum_{N_{1}}\sum_{\beta_{2}}&(Z_{1}-\bar{Z}_{1}(t))^{2}\times P(Z_{1}, N_{1}, \beta_{2}, t)\\
 \sigma^{2}_{NZ}(t)=\sum_{Z_{1}}\sum_{N_{1}}\sum_{\beta_{2}}&(N_{1}-\bar{N}_{1}(t))(Z_{1}-\bar{Z}_{1}(t))\\
 &\times P(Z_{1}, N_{1}, \beta_{2}, t).
\end{split}
\end{flalign}
The probability distribution depends on the potential energy surface (PES) which is defined as:
\begin{flalign}\label{PES}
\begin{aligned}
  U\left(Z_{1}, N_{1}, \beta_{2}, J, R_{\text {cont}}\right)&= \Delta\left(Z_{1}, N_{1}\right)+\Delta\left(Z_{2}, N_{2}\right) \\
  &+V_{\mathrm{cont}}\left(Z_{1}, N_{1}, \beta_{2}, J, R_{\text {cont}}\right)\\
  &+\frac{1}{2} C_{1}\left(\delta \beta_{2}^{1}\right)^{2}
  +\frac{1}{2} C_{2}\left(\delta \beta_{2}^{2}\right)^{2}.
\end{aligned}
\end{flalign}
Here, $\Delta\left(Z_{i}, N_{i}\right)(i=1,2)$ is the mass excess of the fragment, which can be calculated by the liquid drop model. For the shell and pairing corrections terms, we have taken into account the temperature dependence \cite{zhu2018shell}. For the reactions with no potential pockets, $R_{\text {cont}}$ is the position where the nucleon transfer process takes place. $R_{\mathrm{cont}}$ is calculated as $R_{1} + R_{2} + 0.7$ fm. Here, $R_{1,2}=1.16A^{1/3}_{1,2}$. $V_{\mathrm{cont}}$ denotes the effective nucleus-nucleus interaction potential, which contains the Coulomb potential and nuclear potential. The last two terms are dynamical deformation energies of the PLF and TLF. The detailed description of the effective nucleus-nucleus interaction potential $V_{\mathrm{cont}}$ and dynamical deformation energies $\frac{1}{2} C_{i}\left(\delta \beta_{2}^{i}\right)^{2}(i=1,2)$ can be seen in the previous works \cite{PhysRevC.98.034609}.

Assuming the position where the nucleon transfer process takes place is fixed at $R_{\text {cont}}$, the total kinetic energy loss (TKEL) of the primary fragments %configuration $(Z_{1}, N_{1}, \beta_{2})$ formed at the entrance angular momentum $J$ 
can be written as
\begin{flalign}
\begin{split}\label{TKEL}
\text{TKEL} = E_{\text{diss}} + V_{\mathrm{cont}}(Z^{p}, N^{p}, \beta^{p}_{2}, J, R_{\text {cont}})\\
- V_{\mathrm{cont}}(Z_{1}, N_{1}, \beta_{2}, J, R_{\text {cont}}).
\end{split}
\end{flalign}
where the superscript “p” denotes the projectile-target configuration. $E_{\text{diss}}$ is the energy dissipated into the composite system from the incident energy and can be written as
\begin{flalign}
\begin{split}\label{EDISS}
E_{\text{diss}}(J, t) = E_{\text{c.m.}} - V_{\mathrm{cont}}(Z^{p}, N^{p}, \beta^{p}_{2}, J, R_{\text {cont}})\\
- \frac{(J^{\prime}(t)\hbar)^{2}}{2 \zeta_{\text {rel}}}-E_{\text{rad}}(J, t).
\end{split}
\end{flalign}
where $J^{\prime}$(t) is the relative angular momentum that gradually decreases exponentially over time $t$.
$E_{\text{rad}}$($J$,t) is the radial kinetic energy during the collision.

To get the production cross sections of the final products, the code Gemini++ \cite{PhysRevC.82.014610} is employed for the de-excitation process. Subsequent de-excitation cascades of the excited fragments via emission of light particles (neutron, proton, and $\alpha$, etc) and gamma-rays competing with the fission process lead to the final product distribution.

\section{RESULTS AND DISCUSSIONS}\label{sec: results}
We carry out DNS-sysu calculations for $^{136}$Xe + $^{209}$Bi at three above-barriers energyies $E_{\text{c.m.}}=$ 569, 684, and 861 MeV. Note that the experimental conditions for detecting products are imposed \cite{xebi861, xebi526684}: the coverage range of $40^{\circ} < \theta_{\mathrm{c.m.}} < 100^{\circ}$ and 23 MeV $< \mathrm{TKEL}< $ 309 MeV for $E_{\text{c.m.}}=$ 569 MeV, $25^{\circ} < \theta_{\mathrm{c.m.}} < 75^{\circ}$ and 34 MeV $< \mathrm{TKEL}< $ 384 MeV for $E_{\text{c.m.}}=$ 684 MeV and $18^{\circ} < \theta_{\mathrm{c.m.}} < 128^{\circ}$ and 51 MeV $< \mathrm{TKEL}< $ 601 MeV for $E_{\text{c.m.}}=$ 861 MeV. The conditions are taken into account in the following calculations as well.

\begin{figure}[htpb]
    \centering
        \includegraphics[width=8.5cm]{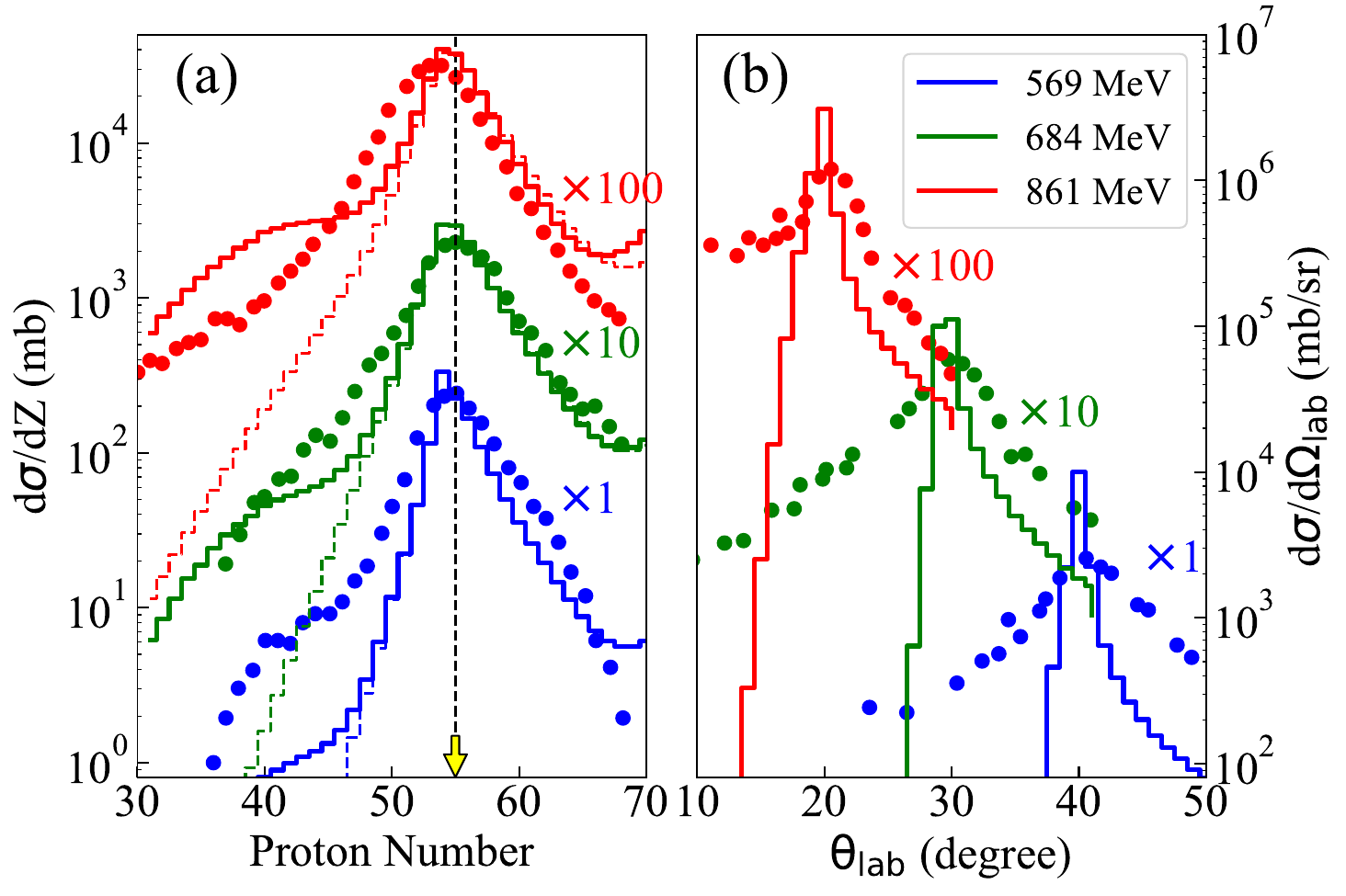}
    \caption{(a) Charge distributions and (b) angular distributions for the $^{136}$Xe + $^{209}$Bi reaction at three collision energies $E_{\text{c.m.}}=$ 569, 684, and 861 MeV. The thick histograms are the calculated distributions of final fragments including the sequential fission events. The black dashed line and the yellow arrow denote the proton number $Z = 55$. The experimental data points (symbols) are taken from Ref. \cite{xebi861, xebi526684}} \label{figure1}
\end{figure}

We proceed with the charge distribution comparison of the experimental data with the model calculations in Fig. \ref{figure1}(a). The charge distributions of the primary fragment are shown by dash histograms, and the final distributions are shown by thick histograms including the sequential fission products. The DNS-sysu model calculations coupled with the Gemini++ code can provide a reasonable description of charge distribution. For light fragments with $Z <$ 50, compared to the primary fragments distributions it can be seen that there is a significant enhancement in final yields due to the fission of excited heavy fragments. However, the contribution from the proton de-excitation process is very weak. As the reaction energy increases, both experimental and theoretical results show that the peak of the charge distribution does not change much, only the variance of the distribution increases. 
%The peak of the experimental charge distribution for $E_{\text{c.m.}}=$ 861 MeV is slightly shifted in the direction of lower charges compared with the DNS calculation. 
%The same theoretical result has been reported by A. V. Karpov and V. V. Saiko for the same reaction based on a Langevin-type approach \cite{PhysRevC.96.024618}. 
The yellow arrow and dashed line in the figure indicate $Z \approx 55$. This suggests that the charge drift velocity should be small for the low TKEL region, in part, owing to the shell effects (cf. the PES of Fig. \ref{figure4}(a)) strongly inhabiting the proton flow. It shows that the shell structure, dissipation, and fluctuation form a complex of interrelated relationship.
In Fig. \ref{figure1}(b), we show the comparison of the experimental angular distributions and the calculations based on the method of calculating the scattering angle proposed in the current model \cite{PhysRevResearch.5.L022021}. The peak positions of the calculations are in good agreement with the experimental values. As the incident energy increases, a very noticeable shift of the angular distribution towards the front angle side is shown clearly.

%At this point, it can be concluded that the present model provides a reasonable agreement with various experimental data for the studied system $^{136}$Xe + $^{209}$Bi at different collision energies. This provides a basis for further analysis of drift and diffusion mechanisms in the MNT reactions.

\subsection{Shell effects on the nucleon drift}
\begin{figure}[htpb]
    \centering
        \includegraphics[width=7cm]{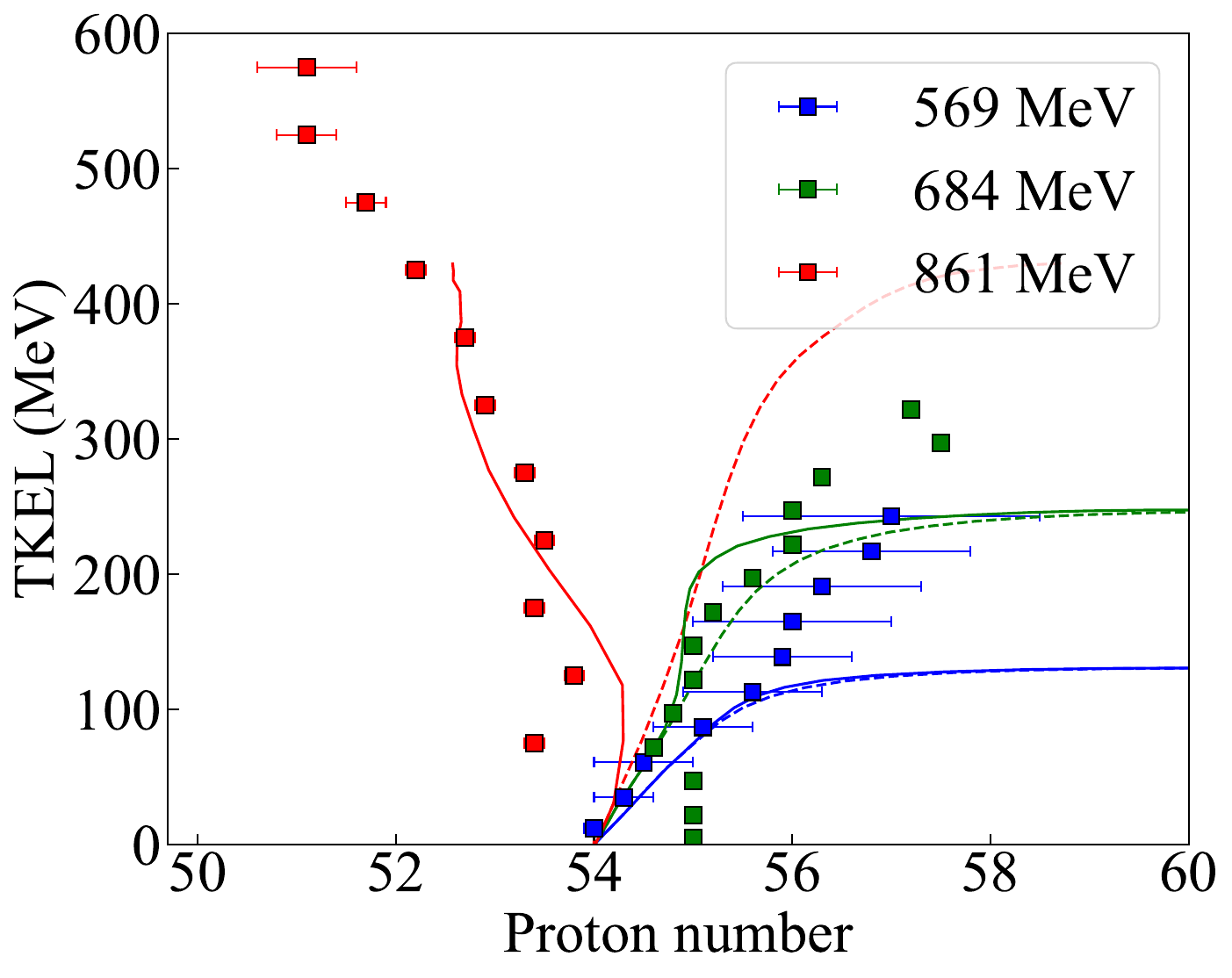}
    \caption{Theoretical (histograms) and measured (symbols) correlation between $\text{TKEL}$ and the averages proton number $\bar{Z}$ of final fragments for the $^{136}$Xe + $^{209}$Bi collision with the $E_{\text{c.m.}}=$ 569, 684, and 861 MeV. The solid line and dashed line denote the final and primary fragments, respectively. } \label{figure2}
\end{figure}

%Drift refers to the systematic motion of particles or entities in a particular direction, typically caused by an external force or gradient. 
%Drift results in nucleons moving in a specific direction, typically from regions of higher potential energy to regions of lower potential energy. 
In the MNT transport process, the nucleon motion could be influenced by the mass or isospin asymmetry of the reaction system \cite{MCINTOSH2019103707, simenel2020timescales, PhysRevC.107.014614}, and even by shell effect \cite{Zagrebaev_2007, zhu2021shell}. 
%In this section, we analyze the drift mechanism for the $^{136}$Xe + $^{209}$Bi reaction at different collision energies $E_{\text{c.m.}}$.
In Fig. \ref{figure2}, we show the correlation between the TKEL and the average proton number $\bar{Z}$ of fragments at $E_{\text{c.m.}} =$ 569 MeV (blue curve), 684 MeV (green curve), and 861 MeV (red curve). Overall, the trend in theoretical calculations aligns with the experimental results. As indicated by the dashed line, the transfer of protons from target Bi to projectile Xe is allowed, favoring the evolution of the dinuclear system toward the symmetric direction due to the positive Q value. When considering the de-excitation process, the drift evolution of fragments is illustrated by the solid line. It can be observed that for low-energy collision at $E_{\text{c.m.}} =$ 569 MeV, the discrepancy between the mean value evolution of the initial and secondary fragments is negligible, which can also be cross-referenced with the above figure (cf. Fig. \ref{figure1}(a)). However, a notable shift in the evolution direction emerges for higher-energy collision $E_{\text{c.m.}} =$ 861 MeV. This is primarily attributed to the contribution of fission products (cf. Fig. \ref{figure1}(a)), and with more intense reactions, the increased fission products further promote the charge evolution of fragment products towards lower charge directions. 
\begin{figure}[htpb]
    \centering
        \includegraphics[width=8.5cm]{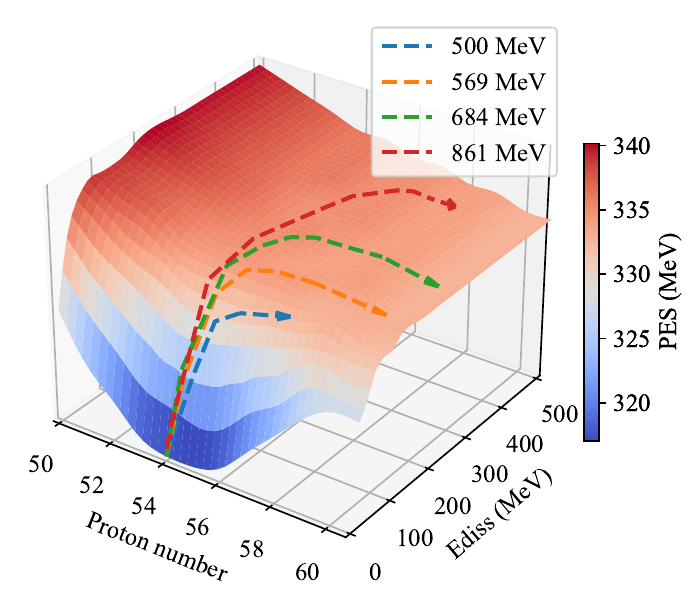}
    \caption{Potential energy surface for the $^{136}$Xe + $^{209}$Bi nuclear system. Different drift evolution paths of the proton number of PLF at the different collision energies are shown schematically by the arrows.} \label{figure3}
\end{figure}

 The calculated PES (cf. Eq. (\ref{PES})) of the nuclear system $^{136}$Xe + $^{209}$Bi is shown in Fig. \ref{figure3} as a function of proton number and the energy dissipated into the system (cf. Eq.(\ref{EDISS})). We can see that as the system heats up, the shell correction of the PES gradually disappears, which to some extent affects the degree of mass and charge drift. For trajectories with different incident energies, the drift trajectory of nucleons on PES will change due to the different energy dissipation rates. 
 
\begin{figure}[htpb]
    \centering
        \includegraphics[width=8.5cm]{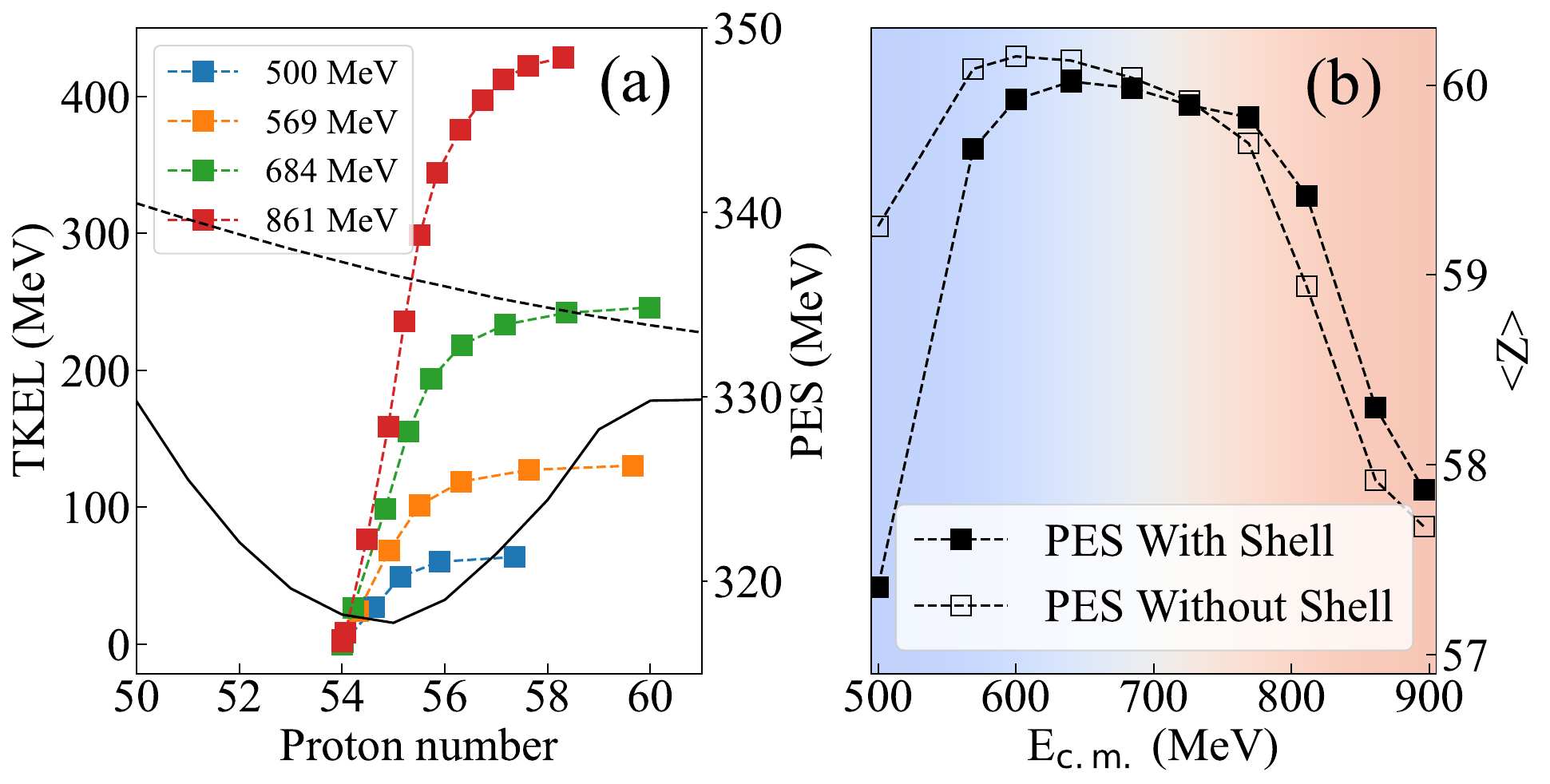}
    \caption{Proton drift evolution of primary fragments for the $^{136}$Xe + $^{209}$Bi reaction at different collision energies $E_{\text{c.m.}}$. In panel (a), the average proton numbers $\bar{Z}$  of the primary PLF as a function of the TKEL. The solid (open) lines denote the result on the potential energy surface with (without) shell correction. In panel (b), the comparison of the average proton numbers $\bar{Z}$ of the primary PLF with the initial angular momentum $J = 50 \hbar$ as a function of the incident energy. The results of with shell corrections and without-shell corrections are shown with solid and hollow black squares, respectively.} \label{figure4}
\end{figure}

 To clarify the influence of the PES on the nucleon drift, the one-dimensional proton PES with (without) shell correction is shown by the solid (dashed) line in Fig. \ref{figure4}(a). The total kinetic energy loss (TKEL) (cf. Eq.(\ref{TKEL})) in the $^{136}$Xe + $^{209}$Bi reaction, are also shown in Fig. \ref{figure4}(a), as a function of the average proton number $\bar{Z}$ of the primary PLF (cf. Eq.(\ref{NZbar})). From the figure, we can see a sharp increase of the $\bar{Z}$ of the primary PLF as the TKEL increases for all incident energies. Then, the system evolves toward the TKEL saturation. The high incident energy enhances the saturation value of TKEL. However, the average proton number $\bar{Z}$ of primary PLF is not positively correlated with the incident energy of the reaction. We can see that with the incident energy increases from 500 MeV to 684 MeV, the maximal value of $\bar{Z}$ increases from 57.3 to 59.9. However, interestingly, for a high incident energy of 861 MeV, the maximal value of $\bar{Z}$ is surprisingly lower than the value in the incident energy of 648 MeV, and a behavior of negative correlation with the incident energy is shown. This reflects that the nucleon drift mechanism in the MNT reaction is not a simple classical thermodynamic phenomenon.

 In Fig. \ref{figure4}(b), the proton values $\bar{Z}$ of the primary PLF with the initial angular momentum $J = 50 \hbar$ as a function of the incident energy is shown by the squares. For the left blue-shaded area, it can be seen that as energy increases, more protons are transferred from the target Bi to the projectile Xe. Conversely, as the energy further increases for the right red-shaded area, the trend of net nucleon transfer will reverse. For the case of the reaction in the red-shaded region, the incident energy is already very high, which may weaken the constraint of PES on the system evolution path. Under an extremely high reaction energy limit assumption, nucleons can disregard the PES and transfer in any direction, exhibiting macroscopic statistical behavior where the average number of protons $\bar{Z}$ remains unchanged. 

By comparing the results of whether PES contains shell correction, only significant differences are observed in the low-energy region. As the energy increases, the influence of the shell effect gradually decreases, indicating the transition of quantum many-body nucleon transfer phenomenon to a classical process (Blue shaded area in Fig. \ref{figure4}(b)). With the continuous increase of energy, the constraint on the evolution path imposed by the PES gradually diminishes, indicating that the nucleon exchange process has transformed into a purely thermodynamic Brownian motion process (Red shaded area in Fig. \ref{figure4}(b)). The combination of these observation results indicates that the quantum shell effect and classical thermal motion jointly influence the proton drift path.

\subsection{Shell effects on the fluctuation}

\begin{figure}[htpb]
    \centering
        \includegraphics[width=7cm]{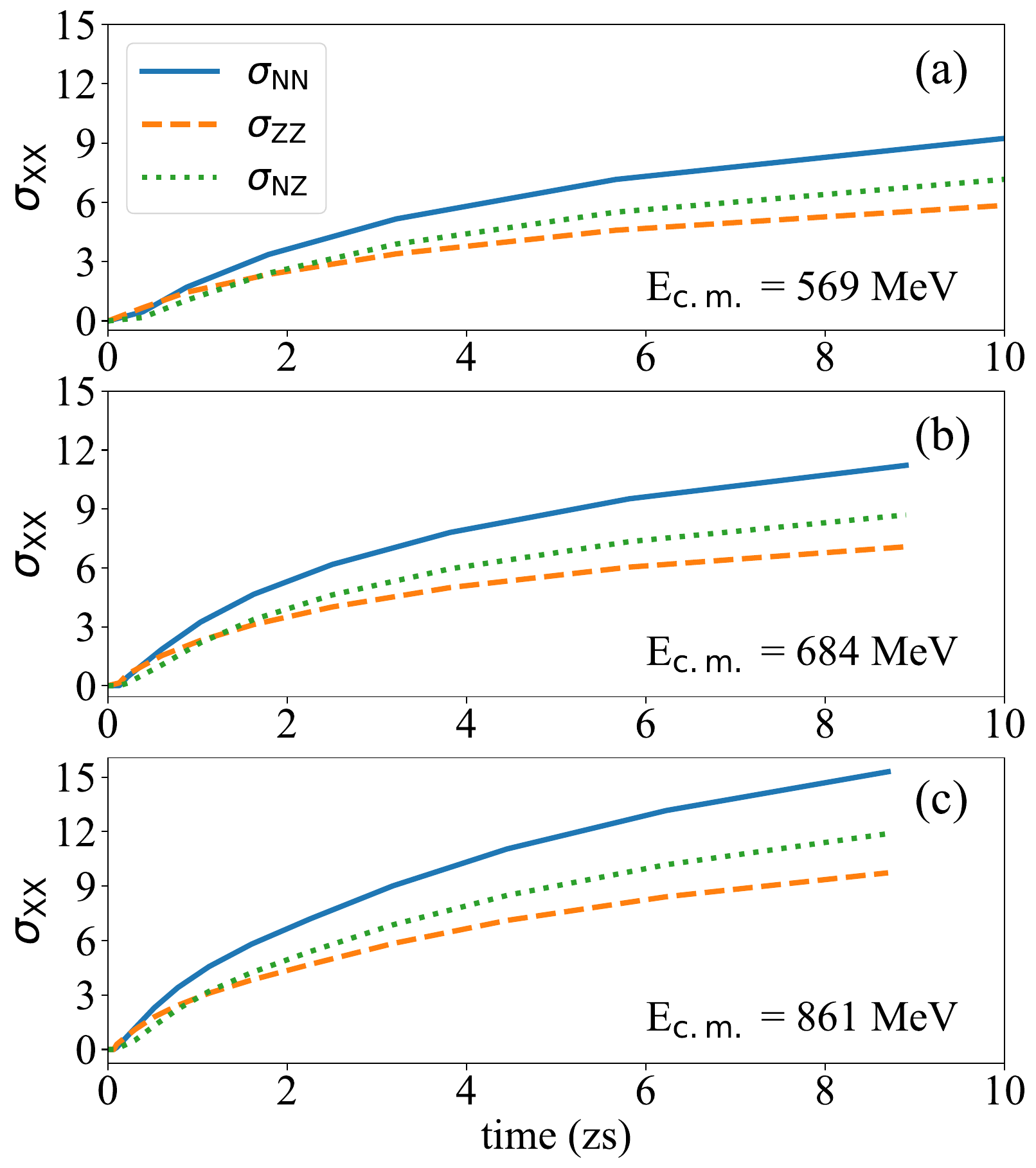}
    \caption{The neutron, proton, and mixed fluctuation as a function of time in the $^{136}$Xe + $^{209}$Bi collisions at the initial orbital angular momentum $J = 50 \hbar$. The panels (a), (b), and (c) denote the calculation with the collision $E_{\text{c.m.}}$ = 569, 684, and 861 MeV, respectively.} \label{figure5}
\end{figure}
% For a classical stochastic process, diffusion is the random motion of particles or entities due to thermal fluctuations. They move in all directions, which leads to a net spread from regions of higher concentration to regions of lower concentration. 
 In the MNT reactions, a significant production of nuclides near the target nucleus has been experimentally observed \cite{watanabe2015pathway, PhysRevLett.130.132502}, which is strongly associated with the fluctuation mechanism. Theoretically, there are many significant efforts in describing the large fluctuation behavior of the MNT reactions \cite{MCINTOSH2019103707, PhysRevLett.105.192701, PhysRevC.79.054606, PhysRevLett.127.222501}. 
 %In this section, we analyze the fluctuation mechanism for the $^{136}$Xe + $^{209}$Bi reaction at different collision energies $E_{\text{c.m.}}$. 
As a direct example, Fig. \ref{figure5} shows neutron, proton, and mixed fluctuation $\sigma_{XX}$ as a function of time for the $^{136}$Xe + $^{209}$Bi collisions at the initial orbital angular momentum $J = 50 \hbar$. The results corresponding to incident energies 569 MeV, 684 MeV, and 861MeV are shown in the Figs. \ref{figure5}(a), (b), and (c), respectively. The temporal evolutions of the variances are obtained by integrating the time-dependent probability distribution $P(Z_{1}, N_{1}, \beta_{2}, J= 50 \hbar, t)$, (cf. Eqs. (\ref{sigma})). From the figures, it is obvious that the relative magnitude orders relationship of various dispersions is similar under different incident energies. In the initial stage of the reaction ($t \lesssim$ 1 zs), we can see the magnitude orders as $\sigma_{NZ} < \sigma_{ZZ} < \sigma_{NN}$. As the reaction continues, we can see that the relationship has changed as $\sigma_{ZZ} < \sigma_{NZ} < \sigma_{NN}$. A similar theoretical result has been reported by K. Sekizawa and S. Ayik in the reaction $^{58, 64}$Ni + $^{208}$Pb within the framework of the microscopic stochastic mean-field approach \cite{PhysRevC.102.014620, PhysRevC.108.064604, PhysRevC.108.054605}. 

From the perspective of classical fluctuation-dissipation theory, for reactions with higher incident energy, as the energy dissipated into the system increases, the fluctuation of the system will also increase. For the Figs. \ref{figure5}(a) to (c), it is obvious that the dispersion value gradually increases with the increase of incident energy. 

\begin{figure}[htpb]
    \centering
        \includegraphics[width=8.5cm]{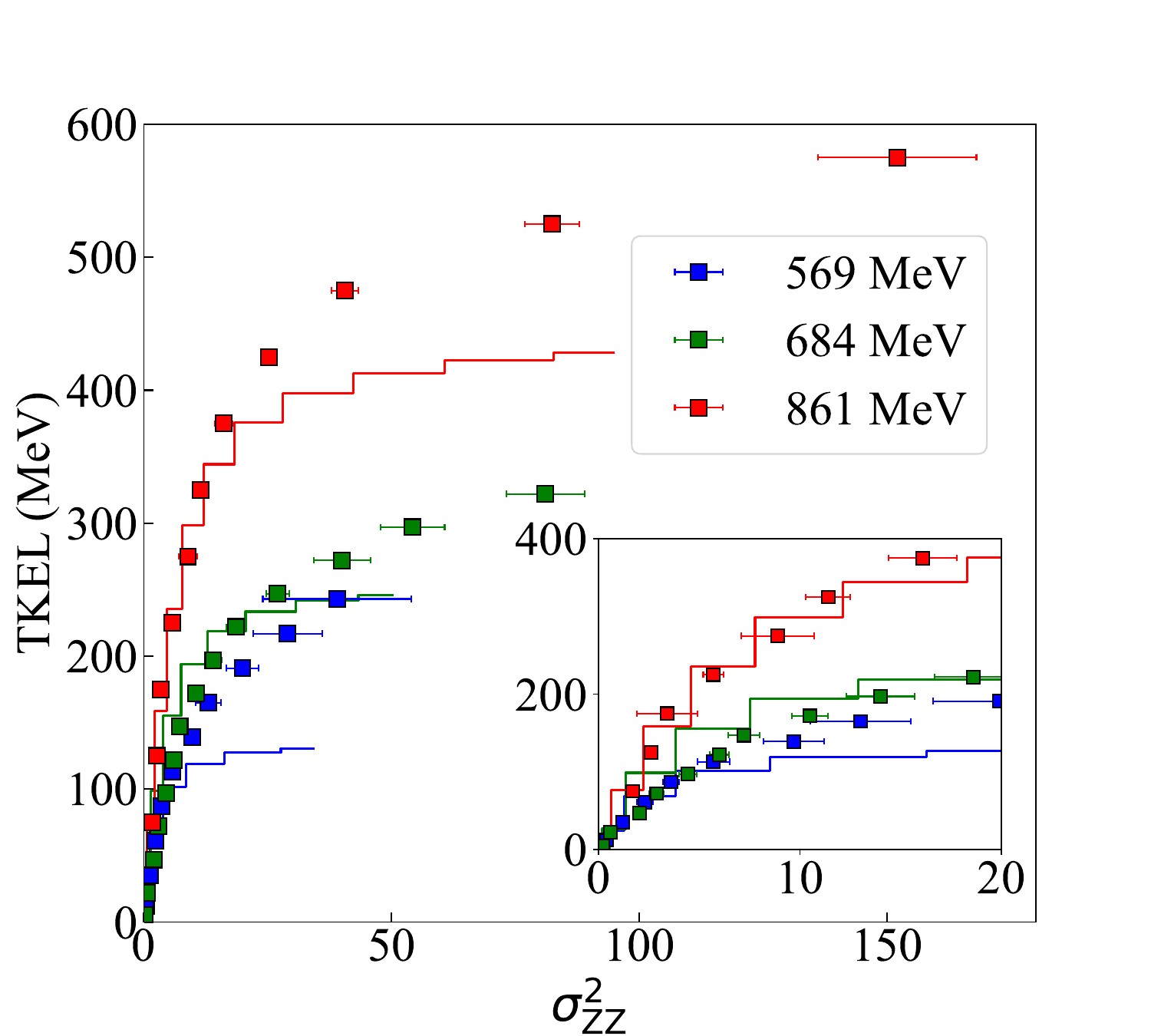}
    \caption{Comparision of calculated (histograms) and measured (symbols) correlation between $\text{TKEL}$ and the variance $\sigma^{2}_{ZZ}$ of the fragment charge distribution for the $^{136}$Xe + $^{209}$Bi collision with the $E_{\text{c.m.}}=$ 569, 684, and 861 MeV. The insert shows the same data on a larger scale. The experiment data is taken from \cite{xebi861}.} \label{figure6}
\end{figure}

The relationship between the variance $\sigma^{2}_{ZZ}$ and the energy loss $\text{TKEL}$ is plotted in Fig. \ref{figure6} for the three bombarding energies. The experiment data points are taken from Ref \cite{xebi861}. The blue, green, and red lines denote the calculation (cf. Eqs. (\ref{sigma}, \ref{TKEL})) with the collision at $E_{\text{c.m.}}$ = 569, 684, and 861 MeV, respectively. For the initial stage in the collision, we find that the model provides a very good description of the correlation between the variance $\sigma^{2}_{ZZ}$ and the total energy loss $\text{TKEL}$. Compared with experimental data at any incident energy, the model works well as one can see from the larger scale diagram ($\sigma^{2}_{ZZ} <$ 20). However, systematic discrepancies remain between theory and experiment for large total energy loss $\text{TKEL}$ values which are not surprising because of the lack of the neck freedom of the degree in the model.  Due to keeping the model exchanging nucleons at a fixed distance while ignoring large deformations in the exit channel, the TKEL calculated by the model (cf. Eq.( \ref{TKEL})) may be slightly smaller for the damped collision.

\begin{figure}[htpb]
    \centering
        \includegraphics[width=8.5cm]{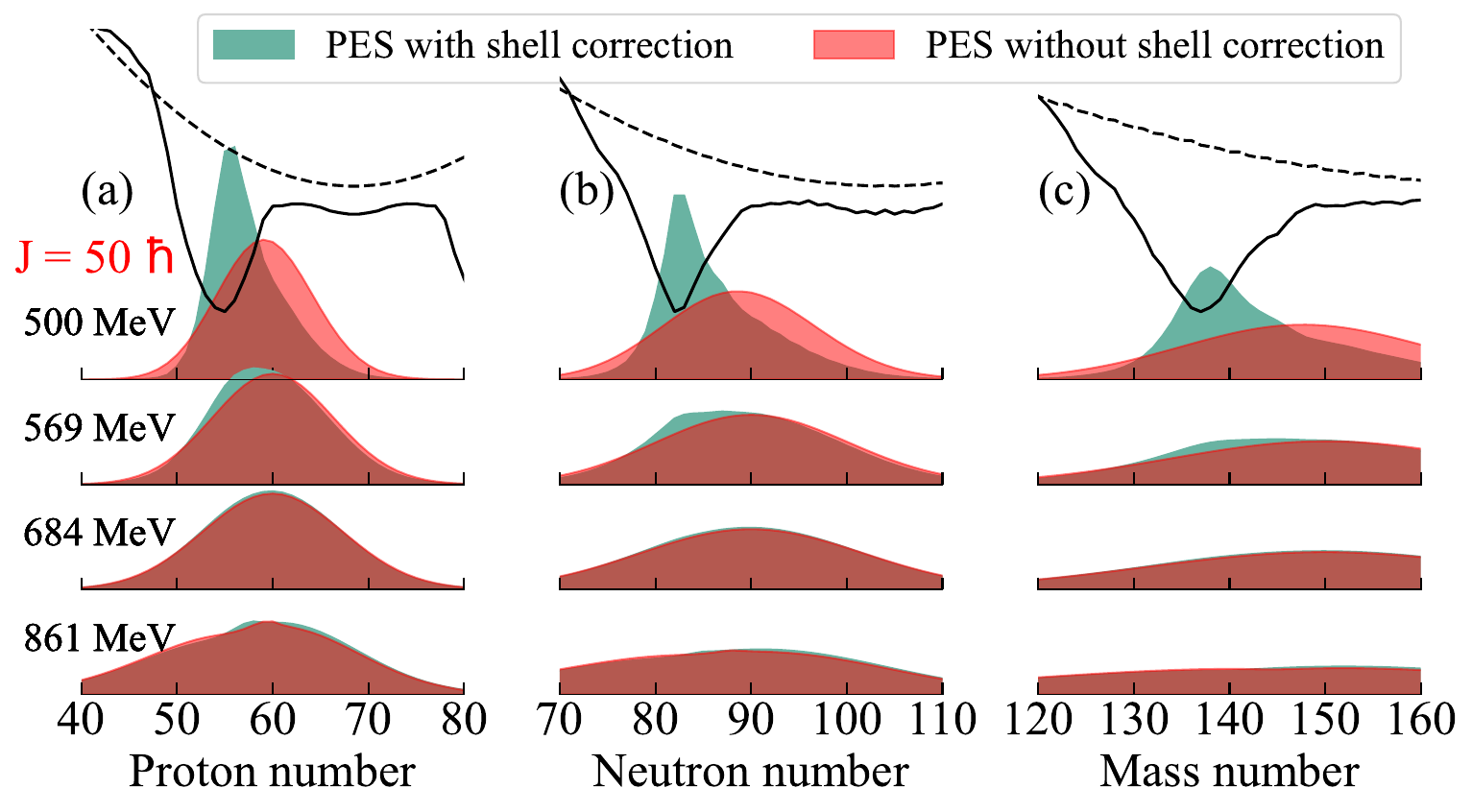}
    \caption{Probability distribution evolution with the initial angular momentum $J = 50 \hbar$ as a function of the incident energy. Panels (a), (b), and (c) represent the probability distribution evolution for the proton, neutron, and mass number of the primary PLF, respectively. The PES of the corresponding dimension with (without shell correction) is represented by a solid (dashed) line.} \label{figure7}
\end{figure}

 As an intuitive representation, we present the probability distributions of the proton number, neutron number, and mass number of PLF with the evolution of incident energy in Figs. \ref{figure7}(a), (b), and (c), respectively. The green patterns represent the results on the PES with shell correction, and the red patterns represent the results on the PES without shell correction. Note that each pattern corresponds to the result with the initial orbital angular momentum $J = 50 \hbar$. From the figures, in addition to the characteristics of the wide distribution formed by the fluctuation mechanism, we can also see the characteristics of mean drift from the peaks of each distribution. These peak positions can also be compared to Fig. \ref{figure4}.

For the collision with the low incident $E_{\text{c.m.}}=$ 500 MeV, it is very obvious that PES with shell correction shows a significant impact on the probability distribution of the reaction, both in terms of its width (fluctuation) and peak value (drift). A solid black line represents the one-dimensional potential energies with shell correction, while a dashed black line represents those without shell correction. Due to the shell attraction, from the neutron, proton, and mass distributions, it can be seen that the primary PLF significantly aggregates at the bottom of the potential energy shown by the green patterns. For the PES without shell corrections, more nucleons tend to flow toward the system's symmetric direction, as indicated by the red pattern.

As the incident energy increases, this aggregation effect for the proton, neutron, and mass number of the primary PLF gradually weakens. The discrepancy between the green and blue patterns for the collision energy $E_{\text{c.m.}}=$ 569 MeV is rather small compared with the discrepancy for the collision energy $E_{\text{c.m.}}=$ 500 MeV. With the incident energy continues to rise, the two distributions almost overlap each other, which means that the shell effect almost disappears. This seems to also align with the physical picture of quantum transport transitioning to classical transport, as the drift mechanism depicted in Fig. \ref{figure4}.

\begin{figure}[htpb]
    \centering
        \includegraphics[width=8.5cm]{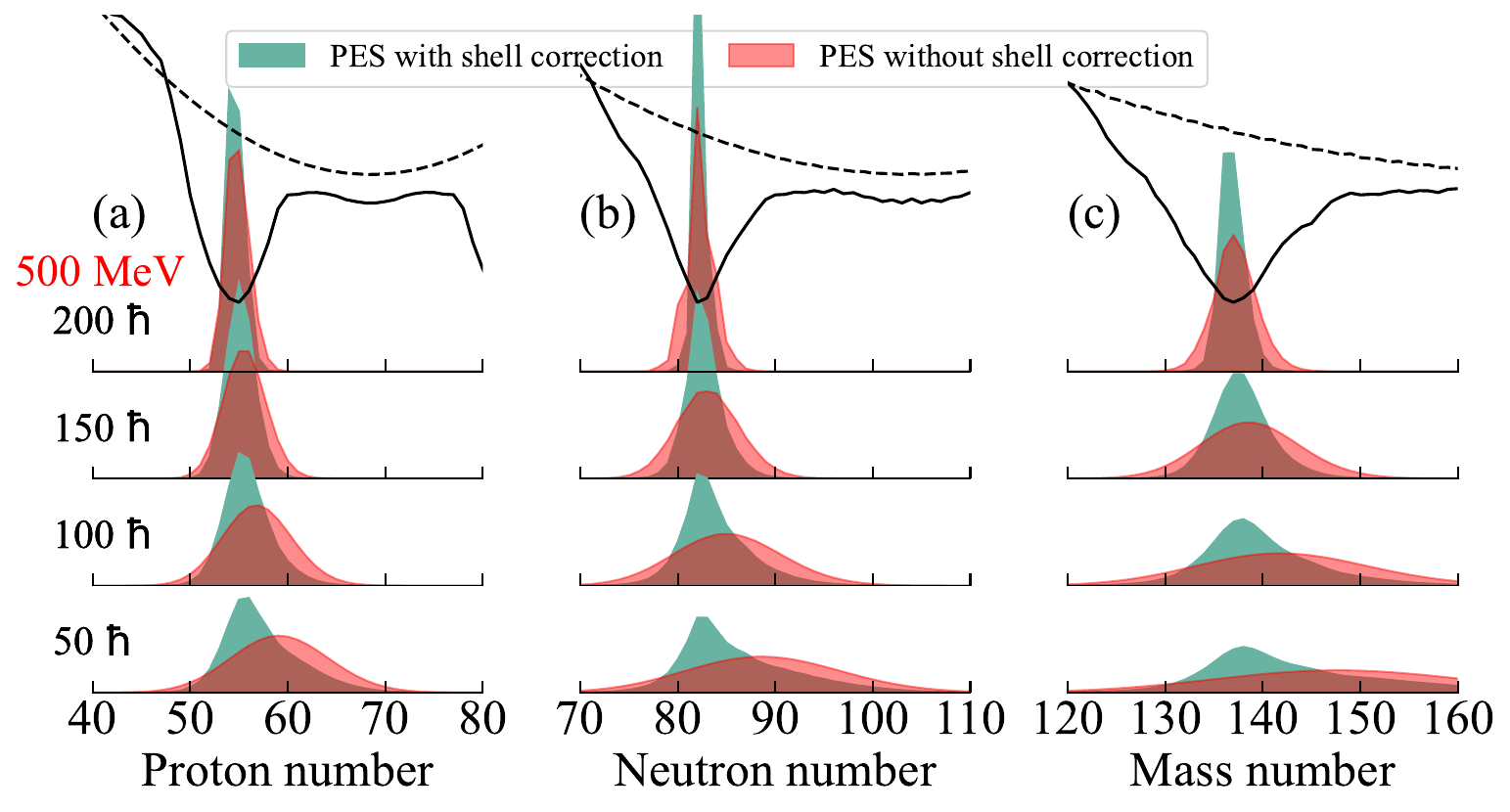}
        \includegraphics[width=8.5cm]{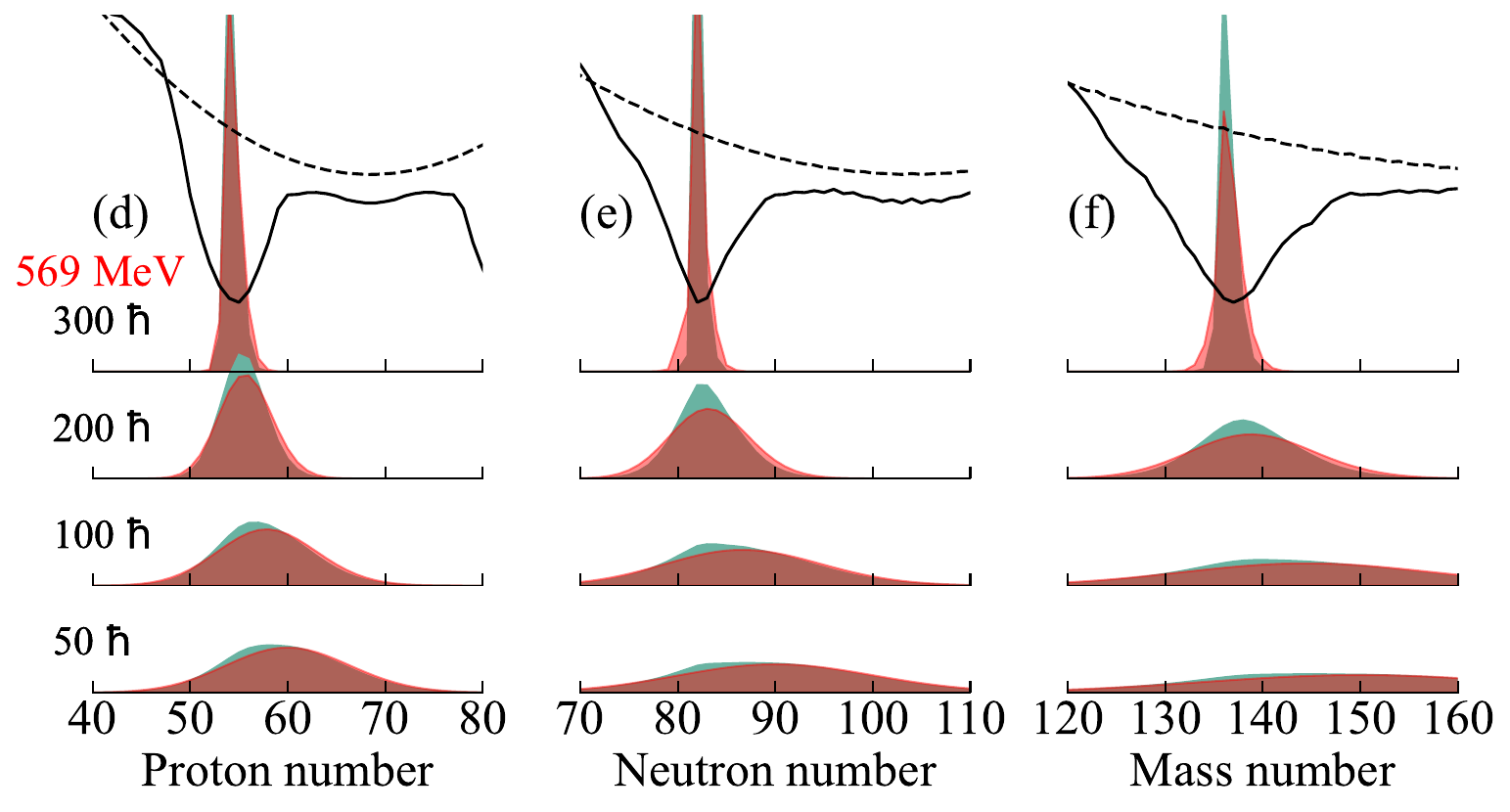}
    \caption{The same as the Fig. \ref{figure7} but for the evolution as a function of the angular momentum.} \label{figure8}
\end{figure}

We also present the evolution of the proton, neutron, and mass probability distributions of the primary PLF for different initial orbital angular momentum at  $E_{\text{c.m.}}=$ 500 MeV and 569 MeV. For the low incident collision energy $E_{\text{c.m.}}=$ 500 MeV shown in Figs. \ref{figure8}(a-c),  the aggregation effect of protons, neutrons, and mass numbers in primary PLF will not show a significant decrease as the angular momentum decreases. The green probability distribution is predominantly concentrated near the bottom of the PES well, while the red probability distribution gradually evolves towards the symmetric region of the PES, with its width expanding gradually. For the case of higher collision energy $E_{\text{c.m.}}=$ 569 MeV, the green and red probability distributions overlap and are almost indistinguishable, even under peripheral collisions ($J = 300 \hbar$). Compared with the result at $E_{\text{c.m.}}=$ 500 MeV, the influence of the shell effect on the nucleon drift and fluctuation would be weak for the high incident energies. 
\section{CONCLUSIONS}\label{sec: conclusions}
In this study, we investigated the drift and fluctuation mechanisms in the collision of $^{136}$Xe + $^{209}$Bi at energies above the Coulomb barrier using the DNS-sysu model. Through a comprehensive comparison between model calculations and experimental data, the DNS-sysu model demonstrated its effectiveness in providing reasonable descriptions of drift and fluctuation dynamics at low energies. 
%Discrepancies between model results and experiments, particularly concerning collective relative motion (total kinetic energy losses, TKEL), prompted further consideration of the radius degree of freedom in the model \cite{ZHU2024138423}.

Analyzing the fragment probability distribution, we calculated the drift evolution path of primary fragments at energies above the barrier. Notably, with increasing system temperature, the shell effect gradually diminishes, leading to increased nucleon transfer as the shell attraction fades. However, at high incident energies, the constraining effect of the potential energy surface (PES) on system evolution weakens, causing a reversal in the evolution direction. In the context of fluctuation, the temporal evolution of neutron, proton, and mass variances of primary fragments at different incident energies was presented. The relative magnitudes of variances underwent a transition during time evolution, and consistent results were obtained using the SMF approach.

Furthermore, the consideration of shell corrections in the PES significantly influenced fragment distribution at low energies, impacting both average values and variances of fragments. However, this influence diminished in high-energy conditions. This comprehensive investigation sheds light on the complex behavior of drift and fluctuation mechanisms in heavy-ion collisions, providing valuable insights into the dynamics of nuclear reactions at various energy regimes. Finally, the anticipated method for reaching the superheavy stable island lies in inverse quasi-fission reactions. In the process, which involves exchanging dozens of nucleons to form a highly excited system, it becomes an intriguing question to explore whether shell effects continue to play a role.

\begin{acknowledgments}
This work was supported by the National Natural Science Foundation of China under Grants No. 12075327; The Open Project of Guangxi Key Laboratory of Nuclear Physics and Nuclear Technology under Grant No. NLK2022-01; Fundamental Research Funds for the Central Universities, Sun Yat-sen University under Grant No. 23lgbj003.
\end{acknowledgments}

% The \nocite command causes all entries in a bibliography to be printed out
% whether or not they are actually referenced in the text. This is appropriate
% for the sample file to show the different styles of references, but authors
% most likely will not want to use it.
%\nocite{*}

%\bibliography{ref}% Produces the bibliography via BibTeX.
%merlin.mbs apsrev4-1.bst 2010-07-25 4.21a (PWD, AO, DPC) hacked
%Control: key (0)
%Control: author (72) initials jnrlst
%Control: editor formatted (1) identically to author
%Control: production of article title (-1) disabled
%Control: page (0) single
%Control: year (1) truncated
%Control: production of eprint (0) enabled
%

\end{document}